# Accurate Measles Rash Detection via Vision Transformer Fine-Tuning

Harshana Rajakaruna[1], Dong Li[2], Anil Shanker[1,3], Qingguo Wang[3]
[1]Office for Research and Innovation, Meharry Medical College, Nashville, TN, USA
[2]Harris School of Public Policy, University of Chicago, Chicago, IL, USA
[3]School of Medicine, Meharry Medical College, Nashville, TN, USA
Corresponding Email: qiwang@mmc.edu

## ABSTRACT
Measles, a highly contagious disease declared eliminated in the United States in 2000 after decades of successful vaccination campaigns, resurged in 2025, with 1,356 confirmed cases reported as of August 5, 2025. Given its rapid spread among susceptible individuals, fast and reliable diagnostic systems are critical for early prevention and containment. In this work, we applied transfer learning to fine-tune a pretrained Data-efficient Image Transformer (DeiT) model for distinguishing measles rashes from other skin conditions. After tuning the classification head on a diverse, curated skin rash image dataset, the DeiT model achieved an average classification accuracy of 95.17%, precision of 95.06%, recall of 95.17%, and an F1-score of 95.03%, demonstrating high effectiveness in accurate measles detection to aid outbreak control. We also compared the DeiT model with a convolutional neural network and discussed the directions for future research.

## 1. INTRODUCTION

The measles virus is among the oldest recorded pathogens known to infect humans [1]. It is highly contagious; up to 90% of susceptible individuals in close contact with an infected person will contract the virus. Moreover, infected individuals can transmit the virus as early as four days before the appearance of the characteristic rash, greatly increasing the potential for spread [2]. In 1912, measles became a nationally notifiable disease in the United States (US), requiring healthcare providers and laboratories to report all diagnosed cases. During the first decade of reporting, an estimated 3 to 4 million Americans contracted measles annually, with approximately 6,000 deaths each year [1].

Following the widespread adoption of vaccination programs, measles was declared eliminated in the US in 2000, with only rare, isolated cases occurring over the next two decades [1]. However, measles resurged in 2025, with 1,356 cases reported across 41 jurisdictions as of August 5, a sharp increase from the 285 cases documented in 33 jurisdictions in 2024 [3]. Most infections were linked to travel-related importations, followed by transmission within under-vaccinated communities [3,4].

Because measles has been exceptionally rare in the US since its elimination, diagnosis has become increasingly challenging, particularly for younger healthcare professionals with limited first-hand experience. The disease's most defining symptom is its characteristic rash, as early signs often mimic other common illnesses. This rash, which typically progresses from the hairline downward across the body, serves as a critical visual cue for clinical diagnosis. Without prompt medical intervention, measles can cause severe complications, including encephalitis, hearing loss, and pneumonia.

Driven by the urgent need for early diagnosis and treatment of skin diseases, such as measles and melanoma, and compounded by the shortage of dermatologists, deep Convolutional Neural Networks (CNNs), a prominent class of deep learning methods, have emerged as the dominant approach for most image-based classification tasks. For instance, Nasr-Esfahani et al.

developed a CNN model for melanoma lesion classification, using segmented skin lesion images as input, and achieved an accuracy of 81% in distinguishing malignant from benign cases [5]. Similarly, Yu et al. applied a deep residual network (ResNet), a CNN architecture, to dermoscopy images, reaching an accuracy of 85.5% [7], while Pham et al. combined data augmentation with the Inception-V4 architecture to achieve 89% accuracy in skin lesion classification [6].

CNNs, however, require large datasets for effective training, a challenge in medical imaging, where acquiring and labeling images demands substantial cost, time, and expert involvement. To address this problem, transfer learning was developed and widely adopted [12–14]. By leveraging knowledge from a pretrained model in a related domain, transfer learning reduces both the amount of data needed for the target task and the training time, while maintaining high performance [8], [9], [16–22]. This strategy was applied in the aforementioned two studies [6, 7], where pretrained ResNet and Inception-V4 were fine-tuned for skin lesion classification. In our recent pilot project, we used transfer learning to develop a deep CNN model, ResNet-50, that achieved an overall accuracy of 95.2%, sensitivity of 81.7%, and specificity of 97.1% for measles rash identification [10]. However, this model was trained almost exclusively on images of individuals with lighter skin tones, limiting its generalizability to those with darker skin. Similarly, Sharma et al. and Ali et al. applied transfer learning to classify various skin conditions and rashes, including measles [11, 26], but failed to report measles-specific performance metrics.

An alternative to CNN-based image classification is the Vision Transformer (ViT), which adapts the Transformer architecture from natural language processing (NLP) to image classification by representing images as sequences of patches [23]. Unlike CNNs, which build global context gradually through localized filters, ViTs use self-attention to capture long-range dependencies from the outset, preserving spatial relationships across the entire image [23, 24]. While ViTs are less data-efficient and typically require large-scale pretraining to match or exceed CNN performance, they often outperform CNNs when such pretraining is available [23]. In medical imaging, ViTs have shown strong results, including melanoma detection with accuracies exceeding other methods [25], making them an attractive candidate for solving the measles rash identification problem across diverse skin tones.

The recent resurgence of measles in the US, coupled with the absence of reliable visual recognition tools, underscores the urgent need for accurate detection tools. In this study, we compiled publicly available measles rash images to fine-tune a pretrained Vision Transformer for measles identification, with the long-term goal of developing an open-source, smartphone-based application to assist both physicians and patients in timely diagnosis. Such a tool could also be integrated into telemedicine platforms and deployed in high-traffic settings, such as airports, to help curb travel-related transmission.

## 2. MATERIALS AND METHODS

### 2.1. Data Collection

In our prior pilot study, we collected measles rash images from the Internet and released them as a publicly available dataset via IEEE DataPort (DOI: 10.21227/9r41-4x79) [10]. The dataset includes rash images from 11 disease states: Bowen's disease, chickenpox, chigger bites, dermatofibroma, eczema, enterovirus, keratosis, measles, psoriasis, ringworm, and scabies, along with images of normal skin. In total, it contains 158 manually curated measles rash images and 1,157 non-measles images, all reviewed by infectious disease specialists. Most images are from individuals with light skin tones.

For ViT model training, we additionally used 755 skin rash images from the publicly available Mpox Skin Lesion Dataset Version 2.0 (MSLD v2.0) on Kaggle [26]. This dataset contains web-scraped images from six classes: mpox (284), chickenpox (75), measles (55), cowpox (66), hand-foot-mouth disease (HFMD, 161), and healthy skin (114) [26]. These images represent 541 unique patients, of whom 30.80% are Black or African American, 29.04% Hispanic or Latina, 27.65% White, 2.66% Asian, and the remainder from other ethnic groups, including

Native Hawaiian [26]. Table I below summarizes the samples used for our model training, comprising the full set of images from both datasets.

**Table 1.** The skin images used for fine-tuning the vision transformer models

| Image class | Number of images |
|---|---|
| Bowen's Disease | 124 |
| Chickenpox | 245 |
| Chigger Bites | 87 |
| Cowpox | 66 |
| Dermatofibroma | 80 |
| Eczema | 95 |
| Enterovirus | 117 |
| Hand-foot-mouth disease | 161 |
| Keratosis | 112 |
| Measles | 212 |
| Mpox | 284 |
| Normal Skin | 155 |
| Psoriasis | 122 |
| Ringworm | 131 |
| Scabies | 79 |
| **Total** | **2070** |

## 2.2. Vision Transformer (ViT) and Baseline Model

We employed the Data-efficient Image Transformer (DeiT) architecture in its base configuration (deit_base_patch16_224: 16×16 patch size and 224×224 input resolution) to identify measles [27]. DeiT is a variant of ViT designed to achieve competitive performance with reduced data requirements by incorporating an efficient training strategy and a distillation token that allows the model to learn from both ground-truth labels and a pretrained teacher network.

In DeiT, each input image of size 224 × 224 × 3 is first divided into non-overlapping patches of 16 × 16 pixels. Each patch is then flattened and linearly projected into a 768-dimensional embedding vector. These patch embeddings, together with a learnable classification token, are passed into a stack of 12 Transformer encoder layers, each comprising multi-head self-attention (12 heads) and feed-forward networks, with residual connections and layer normalization applied at each stage [27]. Positional embeddings are added to preserve spatial information. The output associated with the classification token is fed into a linear classification head to generate the final prediction. Pretrained DeiT weights (model deit_base_patch16_224) were obtained from the Hugging Face model repository via the timm Python library (version 1.0.19) and fine-tuned on the dataset described in Table 1.

To assess DeiT, we also used ResNet-50, a 50-layer residual network, as a baseline, whose pretrained weights were also obtained from Hugging Face. ResNet-50 demonstrated outstanding performance in our pilot project [10] as well as in other related work [11, 27]. Originally developed by Microsoft Research Asia, the ResNet architecture revolutionized deep learning when introduced in 2015, achieving remarkable success across numerous applications and winning multiple image classification competitions, including the ImageNet Large Scale Visual Recognition Challenge (ILSVRC) [18, 19].

In addition to ResNet-50, the ResNet family includes other models such as ResNet-34, ResNet-101, and ResNet-152. In our prior study, however, deeper variants like ResNet-152, although three times deeper than ResNet-50, achieved similar accuracy while requiring substantially longer training times [10]. For this reason, we selected only ResNet-50 as the baseline model for comparison with DeiT. Both models were trained under identical conditions, including learning rate, number of epochs, and batch size, to ensure a fair comparison.

## 3. RESULTS

### 3.1 Fully connected (FC) layer fine-tuning

For both the DeiT and ResNet-50 models, we obtained pretrained weights from Hugging Face using the timm Python library. For each model, all convolutional or transformer encoder layers were frozen, and only the final fully connected (FC) layer was replaced with a new linear layer matching the number of target classes (two in our case). During training, only the parameters of this classification layer were updated, enabling rapid adaptation of the pretrained features to our measles classification task while minimizing computational cost and the risk of overfitting.

Model training was conducted with a learning rate of $3 \times 10^{-4}$, a batch size of 32, and the adaptive learning rate optimizer (Adam) using default $\beta_1$ and $\beta_2$ parameters [28]. Training ran for 30 epochs with early stopping based on validation loss to prevent overfitting. We employed five-fold stratified cross-validation to ensure robustness and generalizability: the dataset was partitioned into five equal folds, with each fold serving once as the validation set (10%) and test set (10%), while the remaining four folds (80%) were used for training. Because some images originated from the same individuals, we used the Python package StratifiedGroupKFold to keep such images within the same fold to avoid data leakage and inflated performance estimates. All experiments were performed in PyTorch (version 2.8.0) with the timm library (version 1.0.19). Reported performance metrics on test data for all five folds are provided in Table 2 and Figure 1.

**Table 2.** Five-fold cross-validation performance of DeiT and ResNet-50 trained on the image dataset listed in Table 1

| Iteration | DeiT performance (%) | | | | ResNet-50 performance (%) | | | |
|---|---|---|---|---|---|---|---|---|
| | Accuracy | Precision | Recall | F1 score | Accuracy | Precision | Recall | F1 score |
| 1 | 95.56 | 95.39 | 95.56 | 95.46 | 91.11 | 91.90 | 91.11 | 87.28 |
| 2 | 97.70 | 97.75 | 97.70 | 97.58 | 89.40 | 90.53 | 89.40 | 85.19 |
| 3 | 93.43 | 92.97 | 93.43 | 92.91 | 88.38 | 78.12 | 88.38 | 82.93 |
| 4 | 95.50 | 95.40 | 95.50 | 95.45 | 91.00 | 91.81 | 91.00 | 87.16 |
| 5 | 93.66 | 93.81 | 93.66 | 93.73 | 90.24 | 81.44 | 90.24 | 85.62 |
| **Average** | 95.17 | 95.06 | 95.17 | 95.03 | 90.03 | 86.76 | 90.03 | 85.64 |
| **Median** | 95.50 | 95.39 | 95.50 | 95.45 | 90.24 | 90.53 | 90.24 | 85.62 |

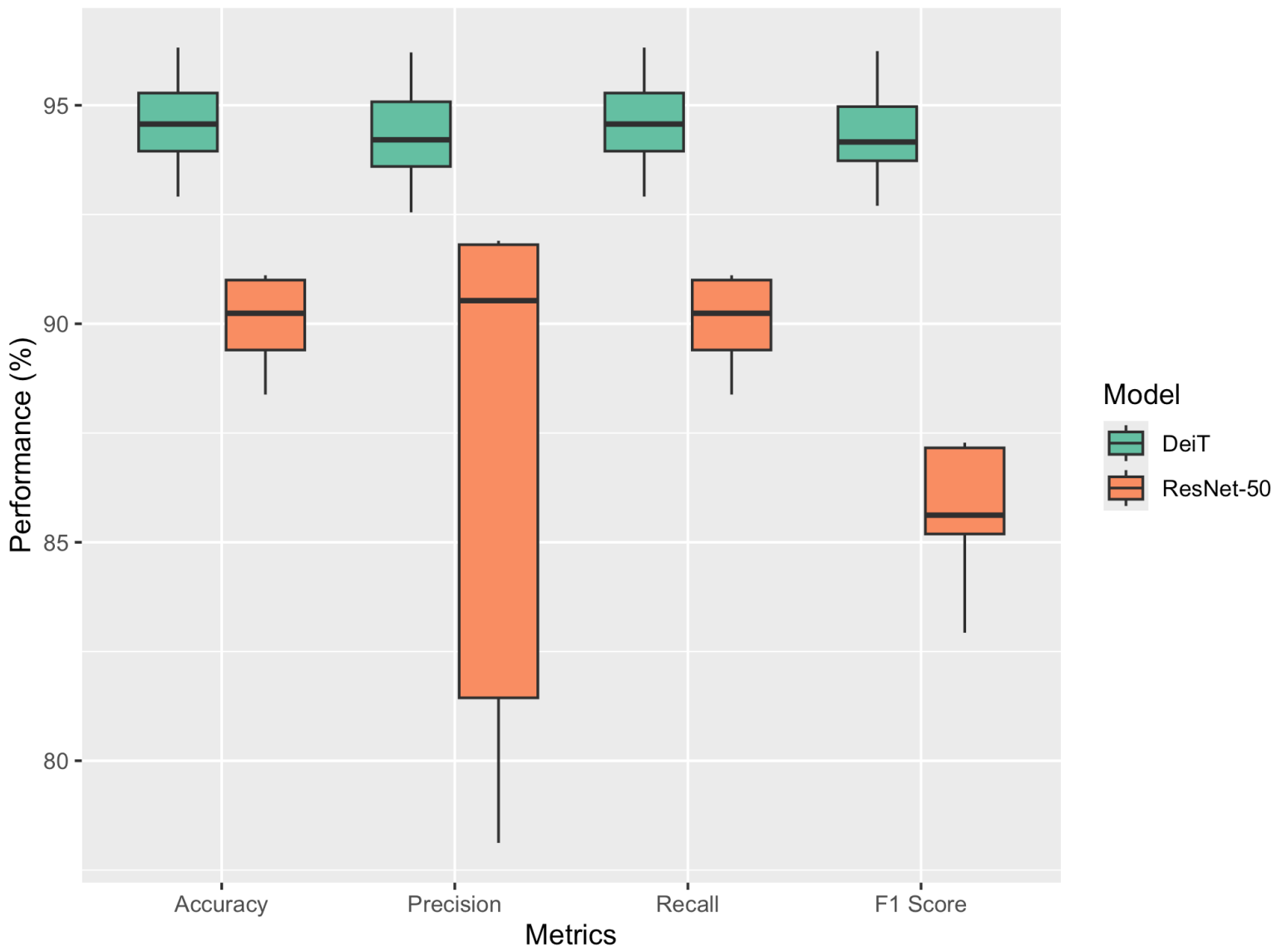

**Figure 1.** Performance comparison of DeiT and ResNet-50 for measles rash identification. The data in the boxplots is from Table 2.

As shown in Table 2, across five independent iterations, the DeiT model consistently outperformed ResNet-50 in all evaluated metrics, including accuracy, precision, recall, and F1 score. DeiT achieved an average accuracy of 95.17%, precision of 95.06%, recall of 95.17%, and F1 score of 95.03%, compared to 90.03%, 86.76%, 90.03%, and 85.64%, respectively, for ResNet-50. The median accuracy for DeiT (95.50%) was also higher than that of ResNet-50 (90.24%), indicating superior and more stable performance across runs. These results demonstrate the advantage of transformer-based models over convolutional neural networks for this classification task when only the FC layer was tuned.

### 3.2 Full model fine-tuning

We also fine-tuned the full DeiT model by adjusting all trainable parameters of the pretrained architecture, rather than updating only the final classification layer. Specifically, we unfroze the patch embedding layer, transformer encoder blocks, and attention mechanisms, and optimized them alongside the classification head during training. This approach enabled the DeiT model to refine low-, mid-, and high-level feature extraction, allowing its learned visual representations to better adapt to the target domain, in this case, measles rash identification.

The five-fold cross-validation results of full model fine-tuning are presented in Table 3, demonstrating consistently high performance in measles rash identification across all evaluation metrics. On average, the model achieved 95.86% accuracy, 95.77% precision, 95.86% recall, and a 95.51% F1-score. While the median values were slightly lower, the differences from the averages were minimal, indicating stable performance. Overall, these results highlight the model's robustness and reliability for accurate measles rash detection.

**Table 3.** Five-fold cross-validation results of full DeiT model fine-tuning

| Iteration | Accuracy (%) | Precision (%) | Recall (%) | F1 (%) |
|---|---|---|---|---|
| 1 | 97.33 | 97.41 | 97.33 | 97.13 |
| 2 | 95.39 | 95.21 | 95.39 | 95.11 |
| 3 | 94.95 | 94.88 | 94.95 | 94.47 |
| 4 | 96.00 | 95.88 | 96.00 | 95.66 |
| 5 | 95.61 | 95.49 | 95.61 | 95.18 |
| **Average** | 95.86 | 95.77 | 95.86 | 95.51 |
| **Median** | 95.61 | 95.49 | 95.61 | 95.18 |

Figure 2 compares the performance of full DeiT model fine-tuning with that of the FC layer tuning. Overall, full model tuning slightly outperformed FC layer tuning. For Precision, the FC layer fine-tuning achieved values between 92.97% and 97.75%, while full model fine-tuning ranged from 94.88% to a peak of 97.41%. Accuracy showed a similar trend, with FC layer tuning ranging from 93.43% to 97.70% compared to 94.95% to 97.33% for full model tuning. Recall mirrored these results, with FC layer tuning scoring between 93.43% and 97.70%, and full model tuning between 94.75% and 97.33%. For the F1 Score, FC layer tuning varied between 92.91% and 97.58%, whereas full model tuning ranged from 94.47% to 97.13%.

The limited performance gain from tuning all layers of the DeiT model, compared to adjusting only its FC layer, may be explained by the small size and imbalance of our measles image dataset. Vision transformers such as DeiT have a large number of parameters, and fully fine-tuning them typically requires substantial amounts of diverse training data to avoid overfitting. Given the relatively few measles rash images available, the pretrained model may have already possessed sufficiently discriminative features, with most of the performance improvement achieved by adapting the classification head to the task-specific distribution. The marginal gains from full fine-tuning suggest that the pretrained feature representations were already well aligned with the dataset's visual patterns.

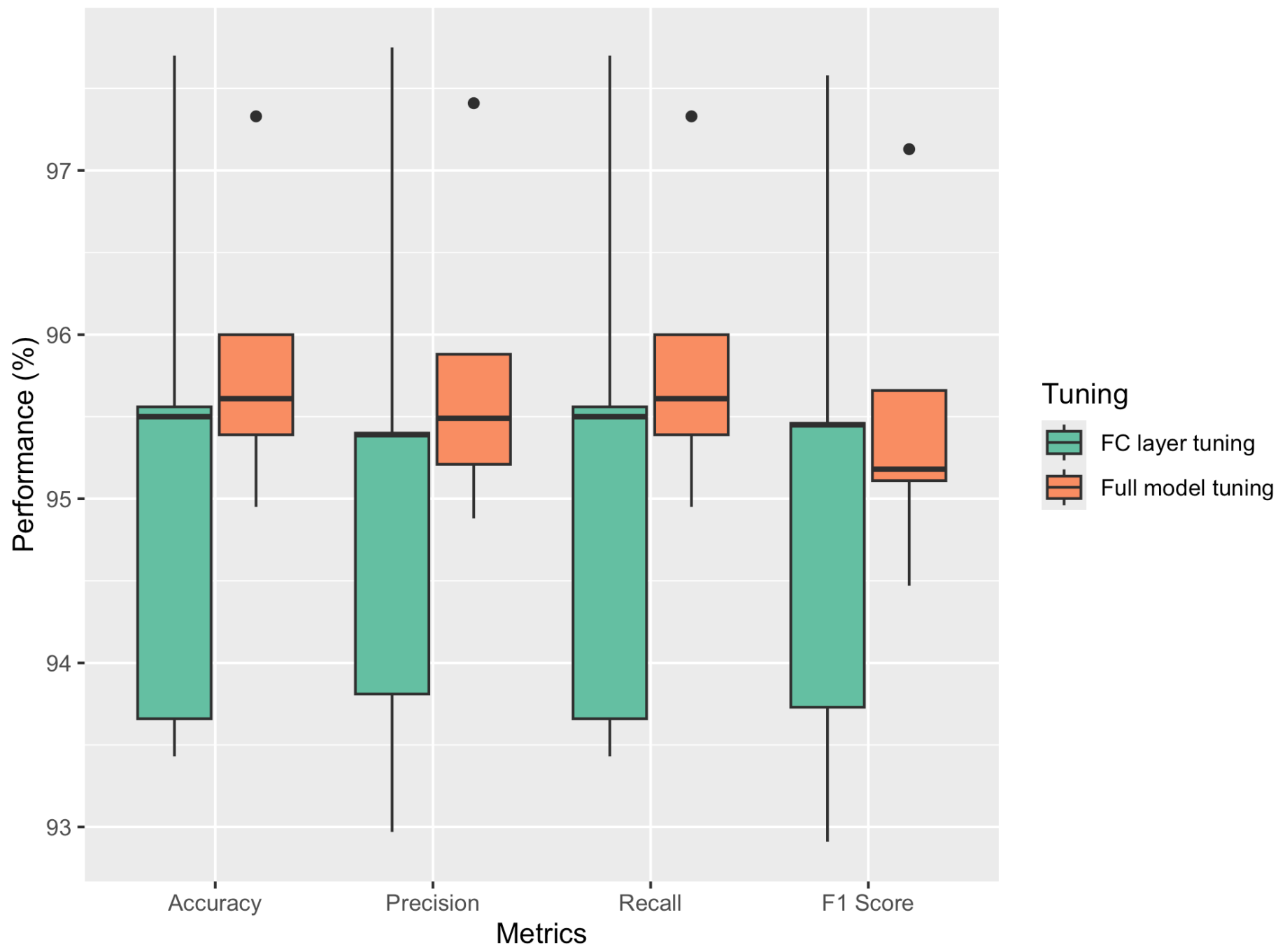

**Figure 2.** Performance of the DeiT model in measles rash identification after FC-layer fine-tuning and full-model fine-tuning. Results for FC-layer tuning are from Table 2, and results for full-model tuning are from Table 3.

## 4. DISCUSSION

With measles resurging in the United States in 2025 and few healthcare providers who can accurately identify it due to its rarity in recent decades, a properly trained model capable of identifying measles rash is essential in combating the outbreak.

In this study, we employed the Data-efficient Image Transformer (DeiT) to distinguish the distinctive measles rash from a range of other skin conditions. After tuning the classification head on our skin rash image dataset, the DeiT model achieved an average classification accuracy of 95.17%, precision of 95.06%, recall of 95.17%, and an F1-score of 95.03%, demonstrating its capability for accurate measles rash detection. For comparison, we also trained a convolutional neural network, ResNet-50, on the same dataset using transfer learning. When the final classification (FC) layer was tuned, ResNet-50 consistently underperformed relative to DeiT, underscoring the advantage of transformer-based architectures for this task.

Our constructed image dataset is predominantly composed of images of Caucasian skin, with comparatively fewer images representing minority skin tones. This imbalance can introduce bias, as the model may misclassify or fail to detect measles accurately in underrepresented groups due to limited exposure to those skin types during training. Addressing dataset imbalance and improving the training process are essential to enhancing the model's accuracy and trustworthiness for underrepresented populations. In addition, the performance of the model can be improved by expanding our training dataset to include a larger spectrum of rash illnesses, e.g., rubella, drug-induced rash, roseola, erythema infectiosum, toxic shock syndrome, Kawasaki disease, as well as multisystem inflammatory syndrome in children, along with others.

Moreover, this study focused solely on the appearance of the rash, without considering its distribution across the body or its progression over time, a notable limitation given that measles rashes characteristically begin on the head or face before spreading downward. Additionally, we did not incorporate information on other concurrent symptoms, such as fever or the classic "3Cs" of measles: cough, coryza, and conjunctivitis. Integrating such clinical and temporal information into the model could further enhance diagnostic accuracy.

## FUNDING

This research was funded by the National Institute of Minority Health Disparities (NIMHD) under grant number U54MD007586, National Institutes of Health (NIH) under Agreement Number 1OT2OD032581, and Chan Zuckerberg Initiative (CZI) grant under award number CZIF2022-007043, and the American Cancer Society (ACS) under grant number DICRIDG-21-071-01-DICRIDG. The views and conclusions contained in this paper are those of the authors and should not be interpreted as representing the official policies, either expressed or implied, of the NIH, CZI, and Meharry Medical College.

## CODE AVAILABILITY

The detailed parameters used and the code for fine-tuning the DeiT model are available at GitHub (https://github.com/qwangmsk/Measles-Detect).

## DISCLOSURES

Conflicts of Interest: The authors declare no conflicts of interest.

## REFERENCE

[1] Centers for Disease Control and Prevention. History of Measles. Available: https://www.cdc.gov/measles/about/history.html. Accessed on August 9, 2025.

[2] Centers for Disease Control and Prevention. How Measles Spreads. Available at: https://www.cdc.gov/measles/causes/?CDC_AAref_Val=https://www.cdc.gov/measles/transmission.html. Accessed on August 9, 2025.

[3] Centers for Disease Control and Prevention. Measles Cases and Outbreaks 2025. Available at: https://www.cdc.gov/measles/data-research/index.html. Accessed on August 9, 2025.

[4] C.I. Paules, H.D. Marston, and A.S. Fauci, "Measles in 2019 - going backward", *NEJM*, vol. 380, no. 28, pp. 2185-2187, 2019

[5] E. Nasr-Esfahani, S. Samavi, N. Karimi, S.M. Soroushmehr, M.H. Jafari, K. Ward, and K. Najarian, "Melanoma detection by analysis of clinical images using convolutional neural network", *In Proc. IEEE Eng. Med. Biol. Soc.,* pp. 1373-1376, August 2016.

[6] T.C. Pham, C.M. Luong, M. Visani, and V.D. Hoang, "Deep CNN and data augmentation for skin lesion classification", *In Asian Conference on Intelligent Information and Database Systems*, Springer Cham., vol. 10752, pp. 573-82, 2018.

[7] L. Yu, H. Chen, Q. Dou, J. Qin, P.A. Heng, "Automated melanoma recognition in dermoscopy images via very deep residual networks", *IEEE Trans Med Imaging*, Vol. 36, no. 4, pp. 994-1004, 2016.

[8] S.S. Yadav, S.M. Jadhav, "Deep convolutional neural network based medical image classification for disease diagnosis", *J. Big Data*, vol. 6, pp. 113, 2019

[9] D.S. Kermany, M. Goldbaum, W. Cai, C.C. Valentim, H. Liang, S.L. Baxter, A. McKeown, G. Yang, X. Wu, F. Yan, J. Dong, "Identifying medical diagnoses and treatable diseases by image-based deep learning". *Cell*, vol 172, no.5, pp. 1122-31, 2018.

[10] K. Glock, et al., "Measles Rash Identification Using Transfer Learning and Deep Convolutional Neural Networks". *Proc. of IEEE International Conference on Big Data (Big Data)*, pp. 3905-3910, 2021.

[11] S. Sharma, "DermaDetect: A novel computer vision model for an accurate diagnosis of skin conditions and rashes", *Proc. of IEEE 7th Int. conf. Data Science and Advanced Analytics (DSAA),* pp. 743-44, 2020.

[12] J. West, D. Ventura, and S. Warnick, "Spring research presentation: A theoretical foundation for inductive transfer", Brigham Young University, College of Physical and Mathematical Sciences. 2007.

[13] T.G. Karimpanal and Bouffanais R., "Self-organizing maps for storage and transfer of knowledge in reinforcement learning", *Adaptive Behavior*, vol. 27, pp. 1-16,2019.

[14] S.J. Pan and Q. Yang, "A survey on transfer learning", *IEEE Trans. Knowl. Data Eng.*, vol. 22, no. 10, pp. 1345-59, 2009.

[15] K. Simonyan and A. Zisserman, "Very deep convolutional networks for large-scale image recognition", *In Proc. Int. Conf. Learning Representations (ICLR).* 2015.

[16] A. Mobiny, A. Singh, H. Van Nguyen, "Risk-aware machine learning classifier for skin lesion diagnosis", *J. Clin. Med.,* vol. 8, no. 8, pp. 1241, 2019.

[17] Y. Liu, A. Jain, C. Eng, D.H. Way, K. Lee, P. Bui, K. Kanada, G . de Oliveira Marinho, J. Gallegos, S. Gabriele, V. Gupta, "A deep learning system for differential diagnosis of skin diseases", *Nat. Med.*, vol. 26, pp. 900-8, 2020.

[18] P.M. Burlina, N.J. Joshi, E. Ng, S.D Billings, A.W. Rebman, and J.N. Aucott, "Automated detection of erythema migrans and other confounding skin lesions via deep learning", *Comput. Biol. Med.,* vol. 105, pp. 151-6, 2019.

[19] L. Fei-Fei. "ImageNet: crowdsourcing, benchmarking & other cool things", *CMU VASC Seminar*, 2010.

[20] K.M. Hosny , M.A. Kassem, and M.M. Foaud, "Classification of skin lesions using transfer learning and augmentation with Alex-net", *PloS ONE*, 2019.

[21] M.A. Milton, "Automated skin lesion classification using ensemble of deep neural networks", *In ISIC 2018: Skin lesion analysis towards melanoma detection challenge*. arXiv:1901.10802 [cs.CV] (2018).

[22] N. Gessert, M. Nielsen, M. Shaikh, R. Werner, and A. Schlaefer, "Skin lesion classification using ensembles of multi-resolution EfficientNets with meta data", *MethodsX,* vol. 7, pp. 100864, 2020.

[23] Dosovitskiy, L. Beyer, A. Kolesnikov, et al., "An image is worth 16x16 words: Transformers for image recognition at scale", *Proc. Of the 9th International Conference on Learning Representations (ICLR), 2021.*

[24] M. Raghu, T. Unterthiner, S. Kornblith, et al., "Do Vision Transformers See Like Convolutional Neural Networks?". *arXiv* :2108.08810v2, 2022.

[25] G.M.S. Himel, M.M. Islam, K.A. Al-Aff, et al. "Skin Cancer Segmentation and Classification Using Vision Transformer for Automatic Analysis in Dermatoscopy-Based Noninvasive Digital System." *International Journal of Biomedical Imaging,* vol. 3022192. 2024.

[26] S. N. Ali, M. T. Ahmed, T. Jahan, et al., "Web-based Mpox Skin Lesion Detection System Using State-of-the-art Deep Learning Models Considering Racial Diversity", *Biomedical Signal Processing and Control*, vol. 98, pp. 106742, 2024.

[27] H. Touvron, M. Cord, M. Douze, F. Massa, A. Sablayrolles, and H. Jégou, "Training data-efficient image transformers & distillation through attention," in *Proc. 38th Int. Conf. Machine Learning (ICML)*, vol. 139, PMLR, pp. 10347–10357, Jul. 2021.

[28] D. P. Kingma and J. Ba, "Adam: A method for stochastic optimization," *International Conference on Learning Representations (ICLR)*, 2015.